\documentclass[10pt,aps,pre,amsmath,amsfonts,amssymb,floatfix,superscriptaddress,notitlepage,nofootinbib]{revtex4-1}
\usepackage{epsfig,amsmath,amssymb,amsfonts,graphicx,color,calc,epstopdf}

\usepackage[T1]{fontenc}

\usepackage{xcolor,hyperref}
\hypersetup{
   colorlinks,
   linkcolor={blue!50!black},
   citecolor={blue!50!black},
   urlcolor={blue!80!black}
}

\let\a=\alpha \let\b=\beta  
   \let\k=\kappa
\let\l=\lambda    
\let\s=\sigma \let\t=\tau

 \def\VV{{\cal V}}

 \def\SS{{\cal S}}
  
\def\ZZ{{\cal Z}}

\def\de{\mathrm{d}}

\newcommand{\beq}{\begin{equation}} 
\newcommand{\eeq}{\end{equation}}
\newcommand{\ba}{\begin{eqnarray}}
\newcommand{\ea}{\end{eqnarray}}

\begin{document}

\title{A continuous constraint satisfaction problem \\
for the rigidity transition in confluent tissues}

\author{Pierfrancesco Urbani}
\affiliation{Universit\'e Paris-Saclay, CNRS, CEA, Institut de physique th\'eorique, 91191, Gif-sur-Yvette, France}

 \begin{abstract}
Models of confluent tissues are built out of tessellations of the space (both in two and three dimensions) in which the cost function 
is constructed in such a way that individual cells try to optimize their volume and surface in order to reach a target shape. 
At zero temperature, many of these models exhibit a rigidity transition that separates two phases: a liquid phase and a solid (glassy) phase.
This phenomenology is now well established but the theoretical understanding is still not complete.

In this work we {consider} an exactly soluble mean field model for the rigidity transition which is based on an abstract mapping. We replace volume and surface functions by random non-linear functions of a large number of degrees of freedom forced to be on a compact phase space. We then seek for a configuration of the degrees of freedom such that these random non-linear functions all attain the same value. This target value is a control parameter and plays the role of the target cell shape in biological tissue models. 
Therefore we map the microscopic models of cells to a random continuous constraint satisfaction problem (CCSP) with \emph{equality} constraints.

{
We argue that at zero temperature, the rigidity transition corresponds to the satisfiability transition of the problem. We also characterize both the satisfiable (SAT) and unsatisfiable (UNSAT) phase. In the SAT phase, before reaching the rigidity transition, the zero temperature SAT landscape undergoes an RSB/ergodicity breaking transition of the same type as the Gardner transition in amorphous solids. By solving the RSB equations we compute the SAT/UNSAT threshold and the critical behavior around it. In the UNSAT phase we also compute the average shape index as a function of the target one and we compare the thermodynamical solution of the model with the results of the numerical greedy minimization of the corresponding cost function. }
\end{abstract}

\maketitle

\section{Introduction}
In recent years there has been a growing interest in understanding the properties of biological tissues.
Tissues can be in different phases: liquid tissues like the blood can (and do) coexist with solid tissues like the epithelial
one. Often, among the same tissue one can see both liquid and solid phases. This is for example the case of tumors inside healthy solid tissues which are often found to have a core-periphery structure \cite{sinha2020spatially, sinha2020self} in which the core has solid-like (or subdiffusive) nature while the periphery is more liquid-like (or superdiffusive). This is crucial to understand metastasis formation and their migration in the organism, {see \cite{blauth2021jamming} for a review}.

Of course the microscopic organization of cells inside individual tissues, changes  a lot depending on the nature of the tissue. The gray and white matter in the brain share very little with how cells are organized in the skin. In the former, the structure of long, up to brain-size, axons and dendrites \cite{kandel2000principles} share more similarities with polymer networks. This structure has been thought to play a role in the gyrification mechanisms both in the cerebral and cerebellar cortex \cite{van20202020, tallinen2016growth}. Packing of compact objects, like colloidal glasses are instead qualitatively closer to epithelial tissues. 
Among different types of tissues, confluent ones have very simple structure. Inside the skin, cells tassellate the space in a rather uniform (yet non-crystalline) manner forming an elastic sheet.

Recently the problem of characterizing the properties of confluent tissues has attracted a lot of interest in the context of active matter \cite{bi2016motility, bi2015density}. Indeed cells can self-propel and migrate and therefore including the effect of this self propulsion may be important to understand their macroscopic behavior.

Simple models of confluent tissues are Vertex {\cite{nagai2001dynamic, yan2019multicellular}} or Voronoi models \cite{bi2016motility}. In both of them, the degrees of freedom realize a convex tessellation of the space\footnote{A convex tessellation is made out of cells that are convex polygons.} (both in two and three dimensions). Given a tessellation, its cost is defined as a sum of surface and volume terms: each cell in the tessellation tries to attain a target volume (area in 2$d$) and surface (perimeter in 2$d$). 
The target volume and surface are not completely independent control parameter: given a target volume it is clear that the target surface cannot be smaller than the one of the sphere (or disk in 2$d$) having that target volume. Conversely, a single cell can have a very small volume while having large surface (think about a thin parallelepiped with one dimension much smaller that the other two). However cells need to tassellate the space and therefore fluctuations to anomalously small volumes are unlikely due to the confluency constraint. 

In order to characterize the behavior of these systems, one can define a target shape index as the adimensional combination of the target surface and volume. 
Recently it has been found that driving the target shape index, one can induce a rigidity transition in these models.
The transition splits the phase diagram into two regions: a liquid phase where cells can diffuse in space, and a solid phase where cells are trapped into an amorphous configuration. Both solids and liquid phases can survive when some driving is switched on (both thermal or active). 

In order to characterize the transition, a series of studies have focused on the landscape of these models at zero temperature \cite{sussman2018no, merkel2018geometrically, merkel2019minimal}. 
In the zero temperature liquid phase, the degrees of freedom manage to find a configuration in which each cell attains the target volume and surface. Instead in the solid phase cells do not manage to achieve their target shape and are stuck in a local minimum of the corresponding energy landscape.

{In order to develop the mean field theory of models of confluent tissues one could consider the infinite dimensional limit as it is done for glasses \cite{SimpleGlasses2020}. This route is sketched in the Appendix \ref{infinite_D}, but it is not pursued since the Voronoi and Vertex constructions are very non-linear and make the analytical derivation very complicated.}
{Instead in this work we consider a simple mean field \emph{model} for the rigidity transition in confluent tissues. }
We take an abstraction step {and instead of deriving the mean field theory of the Voronoi or Vertex models, we consider an abstract model}: seen from a purely optimization point of view, one can think about volume and surface constraints as non-linear \emph{equality} constraints. The minimization of the energy function in tissue models corresponds to finding a configuration of the degrees of freedom such that volume and surface functions are equal to some control parameter. 
Therefore we replace volume and surface functions by non-linear functions. Since the landscape in tissue models is glassy, we will consider random non-linear functions which automatically generate a complex energy landscape. Therefore we map the zero temperature minimization problem to a random constraint satisfaction problem with continuous variables (CCSP) and \emph{equality} constraints.
We will analyze the problem through the replica method and investigate the properties of the SAT and UNSAT landscape and of the zero temperature transition as well.

The manuscript is organized as follows.
{First} we describe the models of two and three dimensional confluent tissues, we review their behavior and map them to random CCSP with equality constraints. Then we solve the CCSP and characterize the liquid (SAT) phase and solid (UNSAT) phase { and the behavior of the model close to the critical point.}

\section{A CCSP for the rigidity transition of confluent tissues}
\subsection{Models of confluent tissues}
Simple models of confluent tissues are Vertex and Voronoi models. To fix the ideas we will focus on two dimensions but everything can be generalized to higher dimensions. In vertex models one has a network of vertices connected by non-intersecting links of varying length. These links define a tessellation of space. The degrees of freedom are the position of the vertices. In Voronoi models instead, one has a fixed number of points, the centers of the cells and construct the Voronoi tessellation out of them. 
For both models the energy function is given in terms of the cells that form the tessellation of the space. Denoting by $A_i$ and $P_i$ the corresponding area and perimeter of the cell $i$, the Hamiltonian of the system is written as
\beq
H[\underline x] = \frac 12 \sum_{i=1}^M \left[k_A(A_i(\underline x) - A_0)^2 + k_P(P_i(\underline x) - P_0)^2\right]
\label{genericH}
\eeq
where we have denoted by $\underline x$ the collective position in phase space of the degrees of freedom and $k_A$ and $k_P$ are elastic constants. We assume that the dimension of the vector $\underline x$ is $N$ which is the system size \footnote{This is the number of vertices or points in Vertex and Voronoi models respectively, multiplied by the dimensionality of the system.}. Instead $M$ is the number of cells. 
The Euler characteristic formula then connects $M$, $N$ and the number of links defining the perimeters of the cells with the genus of the surface that is tessellated. 

In the zero temperature limit one can attain a zero energy $H=0$ if all areas and perimeters match their target values. For the Voronoi model this is not possible in general because $M = N/2$ and therefore one has a number of constraints that is equal to the number of degrees of freedom \cite{sussman2018no} (note that in $2d$ the number of degrees of freedom is $N=2M$). In three dimensions instead the number of degrees of freedom is $N=3M$ which is larger than the number of constraints coming from volume and surface and therefore one can have a genuine liquid phase, where $H=0$ \cite{merkel2018geometrically, damavandi2022energetic}. Note that removing the area term in Eq.~\eqref{genericH} gives rise to a well defined model which can undergo a rigidity transition \cite{sussman2018anomalous}.
Vertex models instead are typically underconstrained since the number of degrees of freedom are typically larger than the number of cells \cite{bi2016motility}. A detailed counting of degrees of freedom and constraints can be found in \cite{merkel2019minimal, damavandi2022energetic}.
In order to induce a rigidity transition one typically consider the setting in which $A_0=A_{TOT}/M$, namely the cells tassellate the space (being $A_{TOT}$ the total area of the system) and change $P_0$. For large values of $P_0$, underconstrained models (namely models for which the number of degrees of freedom is larger than the number of constraints) exhibit a liquid phase ($H=0$). In particular by running some greedy descent on Eq.~\eqref{genericH} it is possible to find a zero energy configuration with an extensive number of zero modes (whose number scales as the number of degrees of freedom subtracted by total number of constraints).
Decreasing $P_0$ one can cross a rigidity transition where greedy search algorithms do not manage to find zero energy configurations.
The density of states in this case does not have zero modes \cite{sussman2018anomalous}.

In order to characterize the critical point one can define the target shape index, a dimensionless control parameter defined as $s_0 = P_0/\sqrt{A_0}$. Furthermore one can also define the average shape index as $s=\langle P\rangle /\sqrt{\langle A\rangle}$ where $\langle P\rangle = \sum_{i=1}^M P_i/M$ and $\langle A\rangle = \sum_{i=1}^M A_i/M$ are the average perimeter and area respectively.
In the liquid phase, at zero temperature, $s_0=s$. Reducing $s_0$, one can cross the critical point at $s_0^*$. Below this point $s$ varies very weakly and stays roughly close to $s^*_0$ \cite{merkel2018geometrically}. The value of $s_0^*$ depends on the geometry of the problem and on the dimension \cite{sussman2018no, merkel2018geometrically, yan2019multicellular}. 

\subsection{The mapping to random CCSPs with equality constraints}
{We are interested in considering a simple model that has the same qualitative properties as the models of confluent tissues just described. We follow the same route that has been recently developed to study the jamming transition in spheres and non-spherical particles \cite{franz2016simplest,FPSUZ17, brito2018universality, ikeda2019mean, FSU19, SU21,sclocchi2022high} }{which we just briefly review. In these cases one is interested in the statistical mechanics of non-overlapping particles. While a mean field theory can be constructed in infinite dimensions, see \cite{SimpleGlasses2020}, one could produce meaningful mean field models by replacing the  constraints of non-overlapping particles with a set of random excluded volume constraints for a set of continuous variables. {The resulting models are abstract CCSPs with \emph{inequality} constraints.} This route has been first suggested in \cite{franz2016simplest}. The models that can be obtained in this way are very effective in describing the physics of $2d$ and $3d$ systems. It is very remarkable the case of jamming of spherical particles where the critical exponents of the transition in the infinite dimensional treatment \cite{CKPUZ14, SimpleGlasses2020} are the very same as the one the random CCSP models \cite{FPSUZ17,FSU19} and agree within numerical precision with numerical simulations in $2d$ and $3d$ systems\cite{charbonneau2021finite, franz2020critical}. For non-spherical particles the models reproduce all features of the jamming/rigidity transition, in particular the complex structure of the density of states \cite{brito2018universality,ikeda2019mean}.}

In order to define a simple mean field model for the rigidity transition in confluent tissues, 
we start by considering an $N$-dimensional vector $\underline x=\{x_1,\ldots,x_N\}$ subjected to a spherical constraint $|\underline x|^2=N$ which defines a compact phase space.
From an optimization point of view the Hamiltonian in Eq.~\eqref{genericH} is a square loss function. We replace the functions $A_i(\underline x)$ and $P_i(\underline x)$ by a set of non-linear functions $h_\mu(\underline x)$.
The Hamiltonian of the model is 
\beq
H[\underline x] = \frac 12 \sum_{\mu=1}^M \left(h_\mu(\underline x)-p_0\right)^2\:.
\label{Hamiltonian}
\eeq
Note that the constraint counting is not the same as in Eq.~\eqref{genericH} since there is no distinction between area and perimeter. However this is not very important since what matters is how the number of constraints $M$ compares with the number of degrees of freedom $N$. 
If the reader wants to have in mind a closer mapping, one can consider the Voronoi model in $2d$ with $k_A=0$. In this case $M=N/2$.
On general grounds we will choose $M<N$ and in particular we will scale $M=\alpha N$ with $\alpha<1$. 

We choose the functions $h_\mu(\underline x)$ to be random non-linear functions. The simplest example is
\beq
h_\mu(\underline x) = \frac 1N\sum_{i<j}J^\mu_{ij}x_ix_j\:.
\label{p2}
\eeq
The components of the $M=\alpha N$ random matrices $J_{ij}^\mu$, with $\mu=1,\ldots,M$ and $i,j=1,\ldots,N$ are i.i.d. random Gaussian variables with zero mean and unit variance. We assume that the matrix is symmetric under index permutations. 
Note that the scaling with the system size $N$ is such that $h_\mu(\underline x)$ is a random quantity of order one.
The control parameter $p_0$ in Eq.~\eqref{Hamiltonian} plays the same role as the target perimeter in the Voronoi model in $2d$ and therefore we will refer to it as the target shape index.

{The model in Eq.~\eqref{Hamiltonian} is rather abstract and does not follow from a microscopic derivation starting from Eq.~\eqref{genericH}. However we will show that: (i) the model has a phase diagram very similar to the one of models of confluent tissues with a liquid phase where all constraints are satisfied and a solid phase where they are all violated; (ii) the behavior of the shape index across the corresponding rigidity transition point has a very similar behavior to the one observed in numerical simulations of models of confluent tissues and (iii) the model is exactly soluble allowing theoretical progress to understand both the nature of the rigidity point, its universality class, the nature of the solid glassy phase and, more broadly, the energy landscape of the model as compared to the one of confluent tissues as discussed in the literature, see \cite{pinto2022hierarchical}.}

The intuitive physics of the model can be summarized as follows. For $\alpha<1$ and sufficiently small $p_0$, there exist, with high probability, configurations $\underline x$ that are at zero energy and therefore such that $h_\mu(\underline x) = p_0$ for all $\mu=1,\ldots, \alpha N$. 
This is because, for random $\underline x$ the variables $h_\mu$ are Gaussian numbers with zero mean and therefore we expect that it is easy to find a configuration with $h_\mu$ nearly zero. In other words a random $\underline x$ is close to be a solution of the CCSP for $p_0\sim 0$. We say that for small values of $p_0$ the model exhibits a zero temperature liquid phase which is satisfiable (SAT). Conversely increasing $p_0$ one is forced to sample the tail of the Gaussian distribution up to the point that finding a zero energy configuration becomes a large deviation. This point is the SAT/UNSAT transition. At large enough $p_0$ the model is in the unsatisfiable (UNSAT) phase and minimizing the Hamiltonian one ends up in a local minimum.

Note that, with respect to confluent tissues where the transition from liquid to solid is crossed \emph{decreasing} $p_0$, here we need to \emph{increase} it (or decrease it to negative values). However this difference is immaterial and depends only on the geometry of the problem.

In the following we will solve the thermodynamics of the model at zero temperature. We underline that a dynamical treatment of the model through dynamical mean field theory { \cite{agoritsas2018out,  Manacorda_2022, mignacco2020dynamical}} is possible and will be developed in a forthcoming work. It will allow to include in the analysis the effect of non-conservative forces (shear drive or activity) and to discuss its differences with thermal noise (whose effect can be instead described by the formalism here below).

\subsubsection{Generalization of the model}
Before discussing the full solution of the model, we mention how it can be generalized to generic non-linear functions $h_\mu(\underline x)$.
Indeed one can generically define $h_\mu(\underline x)$ to be random Gaussian functions with correlations defined by
\beq
\overline{h_\mu(\underline x) h_\mu (\underline y)}=g\left(\frac{\underline x\cdot \underline y}{N}\right)
\label{gen_cov}
\eeq
where we have denoted by the $ \cdot $ the Euclidean scalar product between two vectors and by the overline the average over the realization of these functions.
The special case considered in Eq.~\eqref{p2} corresponds to $g(q)=q^2/2$.
A way to realize the covariances in Eq.~\eqref{gen_cov} is to consider the following form for $h_\mu(\underline x)$
\beq
h_\mu(\underline x) = \sum_{\t=1}^\infty a_\t N^{-\t/2}\sum_{i_1<\ldots < i_\t}J^\mu_{i_1 i_2\ldots i_\t}x_{i_1}x_{i_2} \ldots x_{i_\t} 
\eeq
where $J^{\mu}_{i_1i_2\ldots i_\t}$ are rank $\t$ tensors with Gaussian components with zero mean and unit variance and $a_\t$ are constants. Then, the function $g$ acquires the following Taylor expansion form
\beq
g(q) = a_1^2q+a_2^2 \frac{q^2}2 + \ldots a^2_\t \frac{q^\t}{\t!}+\ldots
\label{mixed_pspin}
\eeq

{
\subsubsection{Related work}} \label{previous_work}
A model of a similar kind of Eq.~\eqref{Hamiltonian} with general covariances as in Eq.~\eqref{gen_cov} was considered in \cite{fyodorov2019spin} as a model for non-linear encryption of a high dimensional signal. However, crucially, in \cite{fyodorov2019spin} only the $\alpha>1$ case was analyzed (this is the regime where signal reconstruction can be information theoretically possible). Furthermore, in \cite{fyodorov2019spin} $p_0$ depends itself on the index $\mu$ and it is correlated with an underlying signal, while here we consider it to be fixed and use it as a control parameter. Having $\alpha<1$ is essential to obtain a model having a rigidity transition at zero temperature.

{
The random model with no signal and $\alpha<1$ was considered and solved in \cite{fyodorov2022optimization} for $g(q)=q$ (see also \cite{ikeda2022sat} for the case $p_0=0$)  and in \cite{tublin2022few} for $g(q)=q^2/2$. In the appendix of \cite{fyodorov2022optimization} the generalization to arbitrary $g(q)$ is also considered. In all these works, however, $p_0$ in Eq.\eqref{Hamiltonian} fluctuates across the different constraints meaning that it is not constant but it is a $\mu$-dependent random variable $p_0\to p_0^\mu$ which is Gaussian with zero mean and unit variance. Moreover, no connection with the physics of confluent tissues is made in these works. 
It is remarked in  \cite{fyodorov2022optimization} that RSB effects, which are absent in the $g(q) \sim q$ case, play a role in the
non-linear case. In \cite{tublin2022few} the RS solution and its instability was obtained and the RSB equations are derived but not solved, leaving open the characterization of the RSB part of the phase diagram and the critical behavior close to the SAT/UNSAT point.
}

{
Here we focus on the model with constant $p_0$ which is useful and relevant for the mapping with tissues. For this model one can discuss the behavior of the average shape index that has been investigated extensively in numerical simulations of the Voronoi model, see for example \cite{merkel2018geometrically}. 
We derive the free energy with the replica method following a standard route paved by previous works \cite{FPSUZ17} and solve the corresponding mean field equations in both the SAT and UNSAT phase to find the SAT/UNSAT transition point. The derivation of the saddle point equations follows also a route that is different with respect to \cite{tublin2022few} which is based on the formulas for the diagonalization of hierarchical matrices \cite{mezard1991replica}. The present derivation is instead useful to compute the behavior of physical observables such as the average shape index. 
The main outcome of the present analysis is:
\begin{itemize}
\item Characterization of the nature of the RSB transition and of the RSB phase for the $g(q)=q^2/2$ model showing that at zero temperature it leads to a continuous fullRSB phase both above and below the SAT/UNSAT threshold.
\item Numerical solution of the saddle point equations in the RSB phase and compution of the SAT/UNSAT transition. This is done by solving the saddle point equations both on the SAT side of the transition and on the UNSAT side. We also compute the behavior of the RSB solution close to the rigidity transition and compare it with other CCSPs.
\item The UNSAT RSB equations are analyzed and the scaling form of the solution in the zero temperature limit is found and solved.
\item Proof that the model with constant $p_0$ that we focus on here has the same phase diagram of the model with Gaussian fluctuating $p_0^\mu$ considered in \cite{fyodorov2022optimization}. The equivalence is found as soon as the fluctuating $p_0^\mu$ are independent Gaussian random variables with zero mean and variance given by $\overline{(p_0^\mu)^2}=p_0^2$ being $p_0$ the control parameter defined in Eq.~\eqref{Hamiltonian}.   Therefore our results in the RSB phase extend also to the case considered in \cite{fyodorov2022optimization, tublin2022few}. This also implies that the RSB transition point derived in \cite{tublin2022few} coincides with the one found here. This is discussed in the Appendix \ref{Equivalence}. However we underline that while these two models share the same phase diagrams, the models of confluent tissues naturally leads to the model with constant $p_0$: indeed  the average shape index can be properly defined only for this model.
\item Comparison of the theoretical results with numerical simulations performed using gradient descent algorithm.
\end{itemize}
}

\section{Solution of the model}

\subsection{The replica approach}
We are interested in studying whether the model has a rigidity transition as a function of $p_0$.
In order to compute the phase diagram we start by looking at the partition function at inverse temperature $\beta=1/T$. We will be interested in the limit $T\to \infty$ and look for the properties of the ground states.
The partition function is defined as
\beq
Z=\int_{|\underline x|^2=N} \de \underline x\, e^{-\beta H[\underline x]}
\eeq
and we are interested in looking at the average free entropy which is given by
\beq
\mathrm f = \frac{1}{N\beta}\overline{\ln Z}\:.
\label{fav}
\eeq
Again, the overline stands for the average over the random couplings which provide a realization of the non-linear functions $h_\mu$.
{In the SAT phase, at zero temperature, $\rm f$ can be used to compute the properties of the zero energy manifold in phase space. Conversely, in the UNSAT phase, $\rm f$ is nothing but minus the ground state energy \cite{FPSUZ17}.}

The average over the disorder in Eq.~\eqref{fav} can be carried on using the replica method. 
Indeed the average free entropy can be written in terms of the replicated partition function through the famous formula \cite{MPV87}
\beq
\mathrm f = \frac {1}{\beta N}\lim_{n\to 0} \overline{Z^n}\:.
\eeq 
For integer value of $n$ one can write $\overline{Z^n}$ as
\beq
\overline{Z^n} = \int \de Q\, e^{NS[Q]}
\eeq
where the action $S[Q]$ is given by
\beq
S[Q]=\frac 12 \ln \det Q +\alpha \ln \ZZ\:.
\eeq
The partition function $\ZZ$ of the 'impurity' problem is given by
\beq
\ZZ = \left.\exp\left[\frac 12 \sum_{ab=1}^n g(Q_{ab})\frac{\partial^2}{\partial h_a\partial h_b}\right] \prod_{a=1}^n e^{-\frac \b2 (h_a-p_0)^2}\right|_{\underline h=0}\:.
\eeq
This quantity gives the local partition function corresponding to a random function $h$ which is in the bath of all the others (from which the name of 'impurity' problem).

The derivation of these formulas follows straightforwardly standard manipulations, see \cite{FPSUZ17} for a detailed reference on a very similar computation \footnote{{In \cite{FPSUZ17}, the Hamiltonian of the system is written as $H=\frac 12\sum_\mu^M (h_\mu-p_0)^2\theta(p_0-h_\mu)$. Therefore the only difference of the model in Eq.~\eqref{Hamiltonian} and the one considered in \cite{FPSUZ17} is the absence of the Heaviside theta function which sets the nature of the constraints (inequalities instead of equalities).}}.
Note that the impurity problem can be solved exactly due to the fact that it can be transformed into a Gaussian integral whose exact evaluation leads directly to the computation of the determinant of hierarchical matrices which can be treated through standard techniques in replica theory, see the appendix of \cite{mezard1991replica} and \cite{fyodorov2022optimization}. However we prefer to stay within the present formalism since it will be convenient to compute the average shape index across the phase diagram.
In the large $N$ limit, the order parameter is the overlap matrix $Q_{ab}$ which is determined by a saddle point condition. At saddle point we can interpret it as $Q_{ab}=\underline x_a \cdot \underline x_b/N$ which is the overlap between two replicas extracted with the Boltzmann measure \cite{MPV87}. 
Following \cite{MPV87}, the matrix $Q$ can be parametrized, in the most general replica symmetry breaking (RSB) form, by a function $x(q):[0<q_m,q_M<1]\to[0,1] $ which encodes for the cumulative distribution of the overlap between two replicas extracted from the corresponding Boltzmann measure. The saddle point for the function $x(q)$ can be written in the following way. We have a first non-linear partial differential equation (PDE) \cite{Pa80}
\beq
\begin{cases}
\frac{\partial m(q,h)}{\partial q} = -\frac 12 \frac{\partial g(q)}{\partial q}\left[\frac{\partial^2 m(q,h)}{\partial h^2}+2x(q)m(q,h)\frac{\partial m(q,h)}{\partial h}\right]\\
m(q_M,h)=\frac{\partial}{\partial h}\ln \int_{-\infty}^\infty\frac{1}{\sqrt{2\pi(g(1)-g(q_M))}}e^{-z^2/(2(g(1)-g(q_M)))-\frac 12\beta(h-z)^2}
\end{cases}
\label{bellman}
\eeq 
which should be solved at fixed $x(q)$.
In order to shorten the notation, we indicate with a prime the derivative w.r.t. $h$ and with a dot the derivatives w.r.t. $q$.
Together with Eq.~\eqref{bellman}, one has a linear PDE given by \cite{SD84}
\beq
\begin{cases}
\frac{\partial P(q,h)}{\partial q} = \frac{\dot g(q)}{2}\left[P''(q,h) -2x(q) \left(P(q,h)m(q,h)\right)'\right]\\
P(q_m,h) = \frac{\exp\left[-(h+p_0)^2/(2g(q_m))\right]}{\sqrt{2\pi g(q_m)}}\:.
\end{cases}
\label{FP}
\eeq 
Finally we have an equation for $x(q)$ in terms of the solution of the PDEs above. This equation is written as
\beq
\frac{q_m}{\lambda^2({q_m})} + \int_{q_m}^q\frac{\de p}{\lambda^2(p)} = \alpha \dot g(q) \int_{-\infty}^\infty \de h P(q,h) m^2(q,h)\:.
\label{closure}
\eeq
From the solution of Eq.\eqref{bellman},\eqref{FP} and \eqref{closure} we can find the profile of $x(q)$.
{Following \cite{FPSUZ17}}, it is useful to take two derivatives w.r.t. $q$ of Eq.\eqref{closure}.
The first derivative gives the following equation
\beq
\frac{1}{\l^2(q)} = \alpha\int_{-\infty}^\infty \de h P(q,h) \left[\ddot g(q) m^2(q,h) + \left(\dot g(q)\right)^2 (m'(q,h))^2\right]\:.
\eeq
The second derivative instead gives
{
\beq
x(q) = \frac{\l(q)}{2}\frac{\alpha \int_{-\infty}^\infty \de h P(q,h) \left[\dddot g(q) m^2(q,h)+3 \dot g(q)\ddot g(q) \left(m'(q,h)\right)^2+\left(\dot g(q)\right)^3 \left(m''(q,h)\right)^2\right]}{\lambda^{-2}(q) + \alpha \l(q) \left(\dot g(q)\right)^3\int_{-\infty}^\infty\de h P(q,h) \left(m'(q,h)\right)^3}\:.
\label{bpoint}
\eeq}
In both equations we have defined $\l(q)$ as
\beq
\lambda(q) = 1-q_M+\int_q^{q_M} \de p\, x(p)\:.
\label{def_lambda}
\eeq
We will study the solution of these equations in detail both in the SAT phase and in the UNSAT phase.
It is useful to derive the equation for the average value of $h_\mu$.
Following the same strategy used in \cite{FPSUZ17} we get
\beq
\overline{h} = \overline{\frac 1M \sum_{\mu=1}^M h_\mu(\underline x)} = p_0 - \frac{1}{\beta}\int_{-\infty}^\infty \de h P(q_M,h) m(q_M,h)
\label{formula_shape}
\eeq
This can be interpreted as the average shape index and should be compared with the target one $p_0$.
At zero temperature, in the SAT phase, we expect $p_0=\overline h$ while in the UNSAT phase this will not be true anymore.

Note that for the moment, all the derivation works for any temperature.
Before focusing on the solution of these equations at zero temperature we will focus on a simpler case that
gives rise to a convex optimization/constraint satisfaction problem and that can be solved fully analytically.

\subsection{A simple case: the convex problem}
\label{Sec:convex}
We will consider the case in which $g(q)=q$. {This case was analyzed and solved in \cite{fyodorov2022optimization} when $p_0$ is fluctuating, see also \cite{ikeda2022sat} for the case $p_0=0$. Here we report its analysis for clarity.} This case corresponds to 
\beq
h_\mu(\underline x) = \frac{\xi^\mu\cdot x}{\sqrt N}
\eeq
with $\xi^\mu=\{\xi^\mu_1,\ldots, \xi^\mu_N\}$ a random vectors with Gaussian entries with zero mean and unit variance. 
In this case, the SAT phase corresponds to solve an underconstrained linear system of equations. However, since the vector $\underline x$ is constrained
to be on the hypersphere $|\underline x|^2=N$, a solution to the linear system will exist only up to a critical value $p_0$.
The Hamiltonian is however convex in the degrees of freedom and therefore the landscape of the problem is simple: in the SAT phase it corresponds to a single ergodic region at zero energy while in the UNSAT phase it is made by a single minimum.
This means that the problem is replica symmetric which implies that $q_m=q_M=q$ and there is no $x(q)$ to take into account. The saddle point equation for $q$ in the zero temperature SAT phase is
\beq
\frac{q}{(1-q)^2} = \alpha \frac{q+p_0^2}{(1-q)^2}\:.
\eeq
At the SAT/UNSAT transition point we expect that $q\to 1$ since the space of solutions reduces to zero \cite{FPSUZ17}.
Therefore the critical point of the rigidity transition is given by
\beq
p_0^* = \left(\frac 1\alpha -1\right)^{\frac 12}
\label{rigidity_RS}
\eeq
Note that the RS solution is stable against RSB as can be checked by a de Almeida-Thouless \cite{de1978stability} stability analysis which in this case gives that the solution is stable for $\alpha\leq 1$ in the SAT phase. Note that $\alpha=1$ corresponds also to a {trivial} upper bound for the SAT phase in a system of random linear equations. {We expect that in the UNSAT phase the RS is also exact since the Hamiltonian defining the model is a convex one in the degrees of freedom, see \cite{fyodorov2022optimization}.}
It is also useful to compute the average shape index $\overline h$. This is trivially equal to $p_0$ in the SAT phase as can be checked by inspection using Eq.~\eqref{formula_shape}. In the UNSAT phase instead one needs to compute the solution of the RS equations. In this case, since the system sits in a minimum of the energy landscape the overlap $q$ must behave as $q=1-\chi T$ being $T\to 0$ the temperature see \cite{FPSUZ17}.
The saddle point equation for $\chi$ is given by
\beq
\frac{1}{\chi^2} = \alpha \frac{1+p_0^2}{(1+\chi)^2}
\eeq
and we note that the unjamming transition corresponds to the limit in which $\chi \to \infty$ since this solution needs to match with the one in the SAT phase where $q<1$. This limit gives the correct SAT/UNSAT transition point as in Eq.~\eqref{rigidity_RS}. Equipped with the equation for the order parameter $\chi$ in the UNSAT phase, we can look at the average shape index. This is given by
\beq
\overline h = \begin{cases}
p_0 & p_0\leq p_0^*\\
\frac{p_0\chi}{1+\chi} = \left(\frac{p_0^2}{\alpha (1+p_0^2)}\right)^{\frac 12} & p_0>p_0^*\:.
\end{cases}
\eeq
In the limit in which $p_0\to \infty$ the average shape index goes to a constant $\overline h \to \alpha^{-\frac 12}$.

\subsection{The SAT/liquid phase}
We will now consider the more generic setting in which $g(q)$ is non-linear giving rise a non-convex optimization/constraint satisfaction problem. To fix the ideas we will derive the results for the case $g(q)=q^2/2$ which is the simplest one (also for numerical simulations).
We note that in this case, the Hamiltonian is invariant under spin inversion symmetry meaning that Eq.~\eqref{p2} is unchanged under $x_i\to -x_i$. 
The generic replica symmetric equation for $q_m=q_M=q<1$ is
\beq
\frac{q}{(1-q)^2} = \alpha \dot g(q) \frac{g(q)+p_0^2}{(g(1)-g(q))^2}\:.
\eeq
For $g(q)=q^2/2$ we get
\beq
\frac{q}{(1-q)^2} = 2\alpha q\frac{q^2+2p_0^2}{(1-q^2)^2}\:.
\eeq
This equation admits a solution $q=0$ which indeed respect spin inversion symmetry \cite{MPV87}. 
{Indeed, the value of $q$ can be rewritten as $q=\sum_i^N\overline{\langle x_i^{(a)} \rangle\langle x_i^{(b)}\rangle}/N$ and the brackets indicate thermal averages. Since the Hamiltonian is invariant under ${\mathbb Z}_2$ one has that $q=0$.}
The stability of the replica symmetric solution is controlled by the de Almeida-Thouless condition which reads
{
\beq
\frac{1}{(1-q)^2} \geq \alpha \dot g(q) \frac{1}{(g(1)-g(q))^2}+\alpha \ddot g(q) \frac{g(q)+p_0^2}{(g(1)-g(q))^2}\:.
\eeq}
For the $g(q)=q^2/2$, using the RS solution $q=0$ we get that RSB takes place when
\beq
1=4\alpha p_0^2
\eeq
which implies that the RS solution is stable up for $p_0<p_0^G$ being the RSB transition point given by
\beq
p_0^G=2 \alpha^{-1/2}\:.
\label{AT_point}
\eeq
At the transition point we can compute the breakpoint of the RSB solution \cite{MPV87} from Eq.~\eqref{bpoint} to get that $x = 0$. This implies that the RSB transition is continuous. Whether the ergodicity broken phase is of fullRSB type or finite RSB type must be checked by computing the 'slope' of the overlap profile $q(x)$ (the inverse function of $x(q)$) \cite{SimpleGlasses2020}. However the formalism we are considering includes the case of finite RSB solutions and since we are going to explicitly solve the equations we will not pursue into this Landau analysis close to the critical point \cite{sommers1985parisi}.
{We note also that the instability point found in Eq.~\eqref{AT_point} coincides with the one presented in \cite{tublin2022few} given the equivalence derived in the Appendix \ref{Equivalence}.}

Therefore this analysis shows that the SAT landscape of the problem, namely the zero temperature liquid phase, is composed by two phases: for small $p_0<p_0^G$ we get a unique ergodic component. Crossing $p_0^G$ one gets a spin glass phase \cite{MPV87} similar to the Gardner phase of amorphous solids \cite{KPUZ13, CKPUZ17} in which the ergodicity of the liquid phase breaks down. Indeed the transition point is a continuous replica symmetry breaking transition. In the $g(q)=q^2/2$ case we have the additional spin inversion symmetry which does not typically pertain to the Gardner phase, see the discussion in \cite{BU15}. However we expect that the properties of the landscape will stay qualitative unchanged for $g(q)=q^2/2+\epsilon q$ for $\epsilon \ll 1$ which does not have a spin inversion symmetry and which we expect by continuity argument that has a continuous RSB transition, this time of the same nature of the Gardner phase.
Therefore {the model has a Gardner transition that separates the zero temperature liquid phase into two regions.}
We note that in \cite{pinto2022hierarchical} an analysis of the landscape of Voronoi models, as compared to the Gardner phase found in amorphous solids, was performed providing evidences of a hierarchical landscape structure of local minima in the solid phase beyond the rigidity point. Here we emphasize that the Gardner/RSB transition happens \emph{before} the rigidity transition is crossed, within the liquid phase. {Therefore the model suggests that signature of Gardner physics may be found also in the liquid phase of models of confluent tissues and this could drive further numerical investigations in Voronoi models.}
As we will see, the rigid/solid phase is in a Gardner/RSB phase too and therefore our model provides a possible theoretical understanding for the numerical findings in \cite{pinto2022hierarchical}. 

In order to enter the Gardner phase we need to solve the RSB equations.
Here we will provide a detailed derivation of the solution. 

\begin{figure}
\centering
\includegraphics[width=0.45\textwidth]{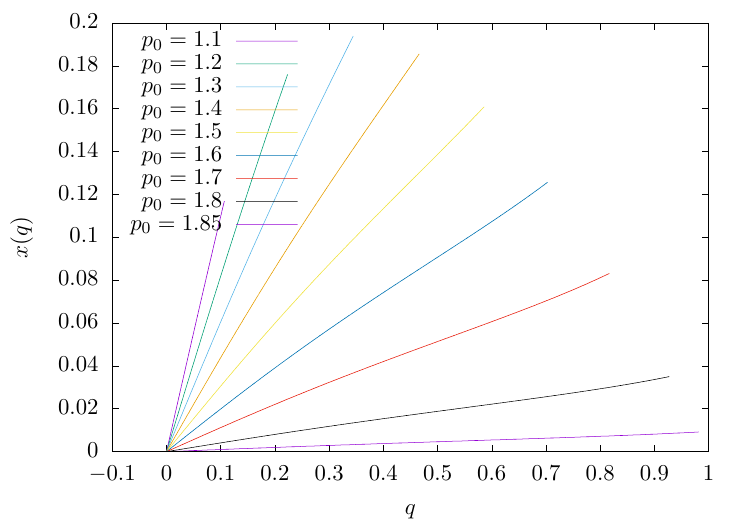}
\includegraphics[width=0.45\textwidth]{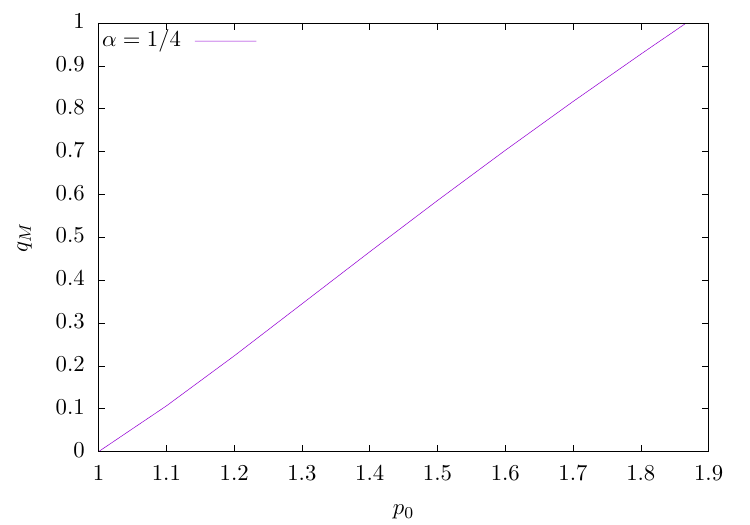}
\caption{Left Panel: the solution for $x(q)$ for some values of $p_0$ in the Gardner phase. The support of this function is in $[0,q_M(p_0)]$. We have realized this plot by solving the RSB equations (see the appendix) for $\alpha=1/4$ and with $N_{\rm RSB}=10000$ steps of replica symmetry breaking. Right Panel: the dependence of $q_M$ as a function of $p_0$.}
\label{Fig_x_SAT}
\end{figure}

The first step towards the RSB solution is to consider the PDEs in Eqs.~\eqref{bellman} and \eqref{FP}. We will show that they can be easily solved.
We start by Eq.~\eqref{bellman}.
In the zero temperature limit, in the SAT phase we expect $q_M<1$ as in general CCSPs, see \cite{FPSUZ17}, and therefore we get
{
\beq
m(q_M,h)=-\frac{h}{g(1)-g(q_M)}
\label{initial_SAT_bellman}
\eeq}
We will now seek for a solution of the corresponding PDE of the form
\beq
m(q,h) = -\frac{h}{R(q)}\:.
\eeq
The function $R(q)$ must be determined by plugging this form into the PDE to get
\beq
\dot R(q) = -x(q) \dot g(q)
\eeq
which, using Eq.~\eqref{initial_SAT_bellman}, can be solved as
\beq
R(q)=g(1)-g(q_M)+\int_q^{q_M}\de p\, x(p) \dot g(p)\:.
\label{def_R}
\eeq
We now turn our attention to Eq.~\eqref{FP}. The boundary condition of the PDE is
\beq
P(q_m,h)= \frac{\exp\left[-(h+p_0)^2/(2g(q_m))\right]}{\sqrt{2\pi g(q_m)}}
\eeq
which suggests to seek for a solution of the form
\beq
P(q,h)=\frac{\exp\left[-(h+S(q))^2/(2D(q))\right]}{\sqrt{2\pi D(q)}}\:.
\eeq
Plugging this ansatz into the PDE we get that
\beq
S(q)=p_0 \frac{R(q)}{R(q_m)}
\eeq
and
\beq
\begin{cases}
\dot D(q)= \dot g(q) \left[1-\frac{2x(q)D(q)}{R(q)}\right]\\
D(q_m)=g(q_m)\:.
\end{cases}
\eeq
The solution of the differential equation for $D(q)$ can be written as
\beq
D(q)=\left[\frac{g(q_m)}{R^2(q_m)}+\int_{q_m}^q \frac{\de p\, \dot g(p)}{R^2(p)}\right]R^2(q)\:.
\eeq 
The final equation for $x(q)$ is given by Eq.~\eqref{closure} which can be written as
\beq
\frac{q_m}{\lambda^2(q_m)} +\int_{q_m}^q\frac{\de p}{\lambda^2(p)} = \alpha \dot g(q) \frac{D(q)+S^2(q)}{R^2(q)}\:.
\label{SP1}
\eeq
Note that this last equation shows that $q_m=0$ for $g(q)=q^2/2$ because of spin inversion symmetry.
It is convenient to write the solution of the equation using directly Eq.~\eqref{bpoint}.
This gives
\beq
x(q)= \frac{\alpha\lambda(q)}{2} \frac{\dddot g(q) (D(q)+S^2(q))+3\dot g(q)\ddot g(q)}{\frac{R^2(q)}{\lambda^{2}(q)}{-}\alpha \lambda(q) \frac{\left(\dot g(q)\right)^3}{R(q)}}\:.
\label{def_x}
\eeq
In the $g(q)=q^2/2$ case we get
\beq
x(q) = \frac 32 \frac{\alpha q \l^3(q)R(q)}{R^3(q)-\alpha q^3 \l^3(q)}\:.
\eeq
We can solve the equation for $x(q)$ in a numerical way as described in the appendix.
We now report the results of the numerical solution of these equations.
We will focus on $\alpha=1/4$ and $g(q)=q^2/2$.
In this case the Gardner transition happens at $p_0^G=1$. 
In fig.\ref{Fig_x_SAT} we report the numerical solution for $x(q)$. 

{First we see that the RSB transition is of continuous (fullRSB) type since $x(q)$ as a support in a continuous interval as we cross the transition.}
We also see that increasing $p_0$ the support of $x(q)$ extends up to $q_M=1$. In the right panel we report the dependence of $q_M$ as a function of $p_0$. We estimate that the SAT/UNSAT transition happens at $p_0\simeq 1.87$. Furthermore we clearly see that $1-q_M\propto p_0^*-p_0$. {Note that this is different from other models, see for example the case of models of jamming of spherical particles, see \cite{FPSUZ17}, where $1-q_M\sim  (p_0^*-p_0)^{\kappa}$ depends on a non-trivial critical exponent $\k>1$ close to the SAT/UNSAT (jamming) transition point. Instead the critical behavior found here is closer to the one of the jamming transition of nonspherical particles \cite{ikeda2019mean}.}

\subsection{The UNSAT/Solid (glassy) phase}
\begin{figure}
\centering
\includegraphics[width=0.45\textwidth]{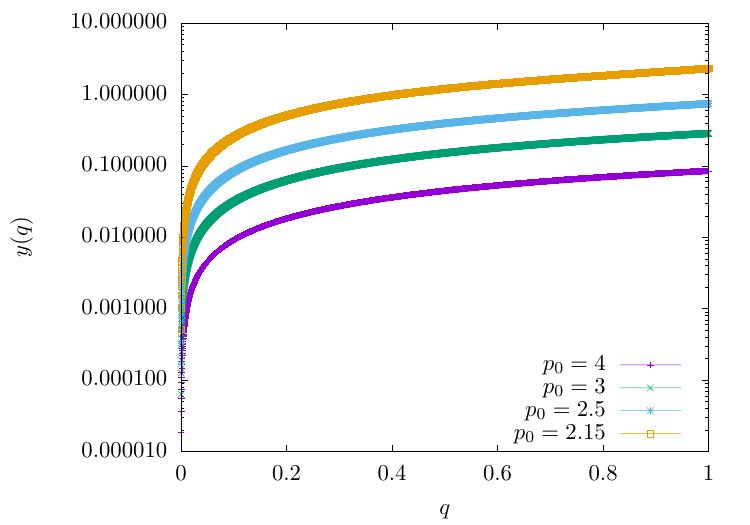}
\includegraphics[width=0.45\textwidth]{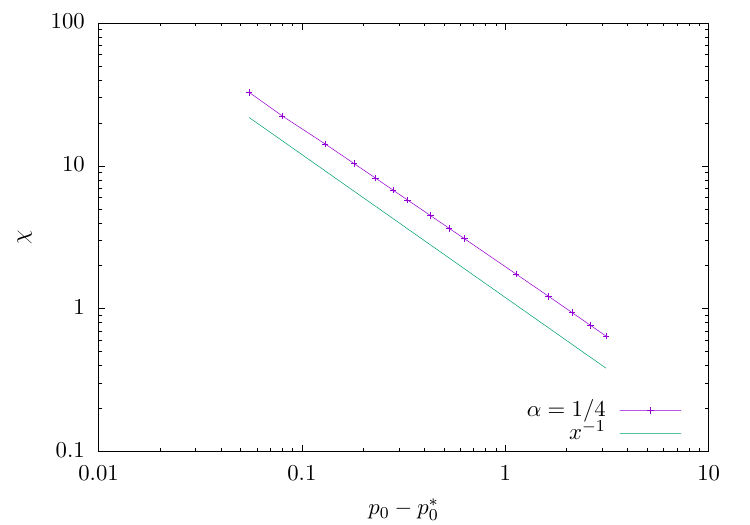}
\caption{Left panel: the profile of $y(q)$ for different values of $p_0>p_0^*$. On approaching the SAT/UNSAT transition $y(q)$ diverges. Right panel: the behavior of $\chi$ as a function of the distance from the SAT/UNSAT transition. The green line is $\sim x^{-1}$ and it is shown only as a guide to the eye.}
\label{fig_yq}
\end{figure}

Having found the rigidity or SAT/UNSAT transition coming from the liquid phase, we turn our attention to the solution of the model in the solid/UNSAT phase.
In this case we look for a scaling solution of the RSB equations in the $T=0$ limit.
It is easy to see by inspection that the saddle point equations admit the following scaling solution
\beq
\begin{split}
q_M&=1-\chi T\\
\hat m (1,h) &= \lim_{T\to 0} Tm(q_M,h) =-\frac{h}{1+\chi \dot g(1)}\\
y(q) &= \lim_{T\to 0} \beta x(q)\:.
\end{split}
\eeq
Plugging this scaling ansatz in the RSB equations we get
\beq
\begin{cases}
\dot{\hat m}(q,h) = -\frac {\dot g(q)}2 \left[\frac{\partial^2 \hat m(q,h)}{\partial h^2}+2y(q)\hat m(q,h) \hat m'(q,h)\right]\\
m(q_M,h)=-\frac{h}{1+\chi \dot g(1)}\:.
\end{cases}
\eeq
Furthermore the equation for $P(q,h)$ becomes
\beq
\begin{cases}
\dot P(q,h) = \frac{\dot g(q)}{2}\left[P''(q,h) -2y(q) \left(P(q,h)\hat m(q,h)\right)'\right]\\
P(q_m,h) = \frac{\exp\left[-(h+p_0)^2/(2g(q_m))\right]}{\sqrt{2\pi g(q_m)}}\:.
\end{cases}
\eeq
We now proceed as before and solve the equations using a simple ansatz of the same form as in the SAT phase. The equation for $\hat m$ can be solved by the following ansatz
\beq
\hat m(q,h) = -\frac{h}{\hat R(q)}
\eeq
with 
\beq
\hat R(q) = 1+\chi \dot g(1) +\int_q^1\de p \dot g(p)y(p)\:.
\eeq
Analogously, for $P(q,h)$ we get
\beq
\begin{split}
P(q,h) &= \frac{\exp\left[-(h+\hat S(q))^2/(2\hat D(q))\right]}{\sqrt{2\pi \hat D(q)}}\\
\hat S(q) &= p_0 \frac{\hat R(q)}{\hat R(0)}\\
\hat D(q) &= \left[\frac{g(0)}{\hat R^2(0)}+\int_{0}^1 \frac{\de p\, \dot g(p)}{\hat R^2(p)}\right]\hat R^2(q)\:.
\end{split}
\eeq
Finally, the equation that determines $\chi$ is
\beq
\int_{0}^1\frac{\de p}{\hat \lambda^2(p)} = \alpha \dot g(1) \frac{\hat D(1)+\hat S^2(1)}{\hat R^2(1)}
\label{SP1_0}
\eeq
where
\beq
\hat \l(q) = \beta \lambda(q) = \chi + \int_q^1 \de p\, y(p)\:.
\eeq
{
Eq.~\eqref{bpoint}, for $g(q)=q^2/2$, in the zero temperature limit gives:
\beq
y(q)=  \frac 32 \frac{\alpha q \hat \l^3(q) \hat R(q)}{\hat R^3(q)-\alpha q^3 \hat \l^3(q)}  
\eeq
which gives
\beq
y(1) = \frac 32\frac{\alpha \chi^3 (1+\chi \dot g(1))}{(1+\chi \dot g(1))^3-\a \chi^3}\:.
\eeq
}
The numerical solutions of these equations gives the profile of $y(q)$ as well as the value of $\chi$.
We will focus again on $g(q)=q^2/2$ and $\alpha=1/4$ so that $q_m=0$.
In Fig.\ref{fig_yq}, left panel, we plot the profile of $y(q)$ for some values of $p_0$ after the SAT/UNSAT transition point.
We clearly see that $y(q)$ diverges on approaching the unjamming or SAT/UNSAT transition from the UNSAT side.
This is consistent with the fact that when we cross the critical point, $x(q)$ must be finite \cite{FPSUZ17}.
The same divergence can be found in the behavior of $\chi$ which must diverge at the critical point since in the UNSAT phase $q_M<1$. 
In the right panel of Fig.\ref{fig_yq} we plot the behavior of $\chi$ as a function of the distance from the SAT/UNSAT transition (whose value, {for a consistency check}, has been estimated by solving the equations on the SAT side and looking at the point where $q_M\to 1$). We see that $\chi\sim (p_0-p_0^*)^{-1}$.

\subsection{The effective shape index as a function of the target one}
\begin{figure}
\centering
\includegraphics[width=0.45\textwidth]{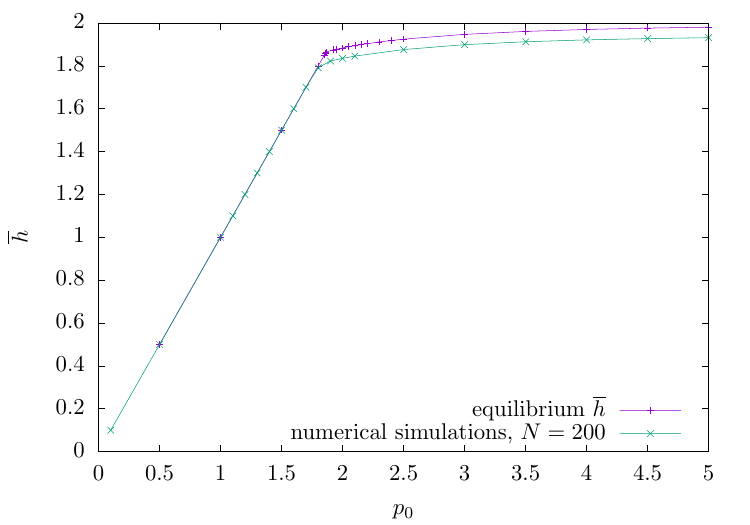}
\caption{The shape index as a function of the of the target one. In green line we have the result of the numerical simulations for $N=200$ and $\alpha=1/4$ for the case $g(q)=q^2/2$. In violet we have the result of the thermodynamic analysis. In principle the two curves are not the same. However we see that the thermodynamic analysis gives a good approximation both for the SAT/UNSAT transition of greedy algorithms and for the average shape index in the solid phase. Whether the SAT/UNSAT transition of the off-equilibrium data coincides with the thermodynamic one must be carefully investigated both from a dynamical and from a landscape point of view.}
\label{shape_fig}
\end{figure}

We conclude the analysis by computing the average shape index as a function of the target one. 
This computation requires to use the solution of the RSB equations in the UNSAT phase.
We get
\beq
\overline h = p_0\left(1 - \frac{1}{\hat R(0)}\right)
\eeq
In Fig.~\eqref{shape_fig} we plot the average shape index as obtained from the RSB solution and compare it with numerical simulations.
{The numerical simulations have been performed with the algorithm described in the Appendix \ref{GD_algo} which is a form of greedy gradient descent dynamics. Therefore the SAT/UNSAT (rigidity) transition obtained from numerical simulations may be, and it is generically expected to be, different from the thermodynamic one. Anyway Fig.~\eqref{shape_fig} shows that the two transition are very close to each other and that the shape index found in the UNSAT phase in local minima of the energy landscape is close to the thermodynamic value.}
A detailed analysis of out of equilibrium dynamics in the large $N$ limit is certainly possible with dynamical mean field theory  textcolor{red}{ \cite{agoritsas2018out,  Manacorda_2022, mignacco2020dynamical}} and will be presented in a forthcoming work.
This will be also important to establish how much the thermodynamic SAT/UNSAT rigidity transition is close to the one of out-of-equilibrium greedy search algorithms.
{In the UNSAT phase a detailed analysis of the landscape of local minima should be also possible due to the intrinsic Gaussianity of the model. This will be interesting as compared to the dynamical treatment and will be the subject of forthcoming works}.

\section{Discussion and Conclusions}
We considered a simple mean field model for the zero temperature rigidity transition in confluent tissues.
The model is based on a mapping of the optimization problem of minimizing the cost function of generic Vertex/Voronoi models
to a random continuous constraint satisfaction problem with inequality constraints. The liquid phase at zero temperature of tissue models corresponds to the SAT phase of the CCSP. Before the SAT/UNSAT transition, corresponding to the rigidity transition in biological tissues, one has a Gardner transition in which the SAT landscape undergoes an ergodicity breaking described by replica symmetry breaking. The transition point shares many similarities with amorphous solids and their Gardner phase.
The occurrence of this phase in real models of tissues must be carefully checked in numerical simulations. 

We note that at finite temperature, the Gardner point gives rice to a full Gardner transition line, which shares the same properties as a spin glass transition. 
Generalizing the model to other forms of $g(q)$ one could provide a phase diagram including RFOT phenomenology which seems to be present at finite temperature far in the UNSAT phase {in the case of confluent tissue models}, see \cite{li2021softness}.
We also underline that the dynamical solution of the model in the large $N$ limit is definitely possible and will be presented elsewhere.
This will also allow to include the effects of driving forces such as shear strain and activity. 
Finally, it would be interesting to develop an analysis of the Hessian of harmonic fluctuation at zero temperature as in \cite{FPUZ15}.
Here the analysis could be carried on by using the technique as developed in \cite{urbani2022field}. We can foresee that the result of this analysis in the 
solid phase will give rise to a gapless density of states following $D(\omega)\sim \omega^\nu$ law with $\nu=2$. This is because the solid phase is marginally stable and therefore the spin glass {susceptibility} is made divergent by {soft linear excitations}\footnote{Note that for the model presented in this model marginal stability is realized at the level of linear excitations.}. This seems different from what observed in \cite{sussman2018anomalous} where a smaller value of $\nu$ seems to be observed (even if the simulations of \cite{sussman2018anomalous} seem to be not sufficient to conclude for any value of $\nu$). However we note that in $d=2$ where the simulations of \cite{sussman2018anomalous} are performed, one expects also the presence of a phononic spectrum which in 2$d$ gives rise to $D(\omega)\sim \omega$. This spectrum may hide the modes corresponding to marginal stability of the underlying energy landscape. Therefore it would be important to repeat the numerical simulations of \cite{sussman2018anomalous} in higher dimensions. 
However, as it happens in the case of structural glasses \cite{lerner2021low}, it could be also possible that the density of states may be affected by quasi-localized excitations in finite dimension whose density deviates from the $\omega^2$ law at finite distance from the rigidity point. It may be possible that the present model could be extended to include such effects on the same lines as in \cite{bouchbinder2021low}.

Finally we would like to underline that the class of constraint satisfaction problems {considered} in this work is rather interesting. The interest is also due to the fact that, while the UNSAT phase depends on the cost function and the square loss considered in this work represents a rather simple case \footnote{We note that introducing higher than quadratic term in the cost function gives rise also to non-linear excitations which may affect the way in which marginal stability is realized}, the SAT phase and the thermodynamic SAT/UNSAT transition is cost-function independent and therefore the analysis presented here is essentially exact. The nature of RSB of the SAT phase is different from other cases, see \cite{FPSUZ17}, where it is much more involved (and richer) due to the emergence of non-linear excitations at the SAT/UNSAT transition point {inherited from the different nature of the constraints}. However the simplicity of the model presented here may be useful to address questions on out-of-equilibrium dynamics which will be the subject of forthcoming works.

{
\paragraph*{Note added -- } After the completion and first submission of this work, the author became aware of the analysis in the appendix of \cite{fyodorov2022optimization} and, especially, the partial results reported in \cite{tublin2022few} where a model very close to the one considered here was analyzed. In both \cite{fyodorov2022optimization, tublin2022few} there is no connection to the physics of confluent tissues which pertains to the present work. The main differences between these works and the present one are reported in Sec.\ref{previous_work}, see also Appendix \ref{Equivalence}.
}

\acknowledgments{I would like to thank D. Dumont and C. Nardini for discussion on a related project which motivated this work and  D. Dumont for a careful numerical verification of some of the results of \cite{damavandi2022energetic}. I would like also to thank R. Guida, F. David, M. Paoluzzi and  V. Ros for useful discussions. This work is supported by 'Investissements d'Avenir' LabEx PALM (ANR-10-LABX-0039-PALM).
}

\appendix
\section{Algorithm for the numerical solution of the RSB equations}
We will give a detailed account on the numerical procedure to solve the RSB equations.
Since this is essentially the same for the SAT and UNSAT phase, we will write explicitly the algorithm only for the SAT phase.
We will also focus on the case $g(q)=q^2/2$. 
We need to find the profile $x(q)$ and the value of $q_M$.
In a nutshell the algorithm works by starting from a guess of $q_M$ and construct $x(q)$ by integrating backwards the equations for $\lambda$ and $R$. From the estimate of $x(q)$ one can recompute a new guess for $q_M$.

The pseudocode is the following.
We discretize the interval $[0,q_M]$ in $L-1$ 'time' steps and define $\de t=q_M/(L-1)$.
Then we go as follows:
\begin{enumerate}
\item Start from a guess for $q_M<1$
\item initialize:
\beq
\begin{split}
\lambda(q_M) &= 1-q_M\\
R(q_M) &= g(1)-g(q_M)\\
x(q_M) &= \frac{2\alpha q_M \l^3(q_M)R(q_M)}{R^3(q_M)-\alpha q_M^3\l^3(q_M)}
\end{split}
\eeq
\item for $i=1$, to $i=L-1$ do
\begin{enumerate}
\item get $\lambda(q_i)$ from Eq.~\eqref{def_lambda}
\item get $R(q_i)$ from Eq.~\eqref{def_R} 
\item get $x(q_i)$ from Eq.~\eqref{def_x} 
\end{enumerate}
\item update $q_M$ by
\beq
\begin{split}
q_M^{\rm new} &=\alpha \dot g(q_M) \frac{D(q_M)+S^2(q_M)}{R^2(q_M)}-\int_{0}^{q_M}\frac{\de p}{\lambda^2(p)} +q_M\\
q_M&\leftarrow\tau q_M^{\rm new} + (1-\tau)q_M
\end{split}
\eeq
\item repeat points 3 and 4 up to convergence
\end{enumerate}
Note that we have introduce a damping factor $\t$ to remove dangerous oscillations that could disturb the convergence to the fixed point. 

\section{Numerical simulations}\label{GD_algo}
Numerical minimizations were performed to produce the data plotted in Fig.\ref{shape_fig}. Here we outline the numerical algorithm. We followed the very same strategy as in \cite{FPUZ15} which we recall here. We used the bfgs2 routine of the GSL library \cite{galassi2002gnu}. 
However since the model is defined with a spherical constraint, we cannot use the bfgs2 routine as it is implemented.
We overcome the problem by define a new Hamiltonian \cite{FPUZ15}
\beq
\hat H(\underline x) = H\left(\frac{\sqrt N \underline x}{|\underline x|}\right) + \frac A4 \left(|\underline x|^2-N\right)^2
\eeq
{The first term of $\hat H$ is automatically computed on a vector with the right length.} The second term instead fixes the norm of the vector $\underline x$ to the desired value. The constant $A$ can be chosen to be of order one.

\section{A comment on infinite dimensional models}\label{infinite_D}
We wanted just to point out that while the model presented in this paper is rather simple and schematic, one could develop a first principle theory of Voronoi models in the large dimensional limit following a route recently explored for structural glasses {\cite{SimpleGlasses2020}}.
In order to clarify this point we consider a set of $N$ points in $d$ dimensions. 
We focus now on Voronoi models. 
For a given point $x_i$ (now a vector in $d$ dimension) the $d$-dimensional volume of the corresponding Voronoi cell can be written as
\beq
\VV_i(\underline x) = \int \de y \prod_{j\neq i} \theta(|x_j-y|-|x_i-y|)
\eeq
It is useful to defined a modified volume as
\beq
\VV_i(\underline x,\sigma) = \int \de y \prod_{j\neq i} \theta(|x_j-y|-|x_i-y|+\sigma)\:.
\eeq
The surface of the corresponding Voronoi cell can be obtained as
\beq
\SS_i(\underline x) =\left. \frac{\partial V_i(\underline x, \s)}{\partial \s}\right|_{\s=0}\:.
\eeq
We note that in infinite dimensions $d\to \infty$, we expect that $V_i\sim e^{d\Sigma}$ and the surface has the same scaling as well.
Therefore one can define the shape index as
\beq
s_i(\underline x) = \frac{\SS_i(\underline x)}{\VV_i(\underline x,0)} = \left.\frac{\partial}{\partial \sigma} \ln \VV_i(\underline x_i,\s)\right|_{\s=0} \sim d
\label{shape_large_d}
\eeq
and consider an Hamiltonian of the form
\beq
H(\underline x) = \frac 1{2d^2} \sum_i^N(s_i(\underline x)-s_0)^2\:.
\label{normal_H_d}
\eeq
The interesting interpretation that comes out from this formulation is that, as it is evident from Eq.~\eqref{shape_large_d}, the shape index $s_i(\underline x)$ is a response function computed as the derivative of a local free energy.
This is also true in the mean field model as it is evident from the definition of the impurity problem and the expression of the average shape index {given in Eq.~\eqref{formula_shape}}.

{
\section{The model with random $p_0^\mu$.} \label{Equivalence}
In \cite{fyodorov2022optimization, tublin2022few}, a different version of the model in Eq.~\eqref{Hamiltonian} was considered. In particular the constant $p_0$ is promoted to fluctuate and therefore it becomes $\mu$-dependent. The random variables $p_0^\mu$ are independent identically distributed Gaussian random variables with zero mean and variance defined as $\overline{(p_0^\mu)^2} = \sigma^2$. The equivalence between the saddle point equations of the model with constant $p_0$ and the one with fluctuating $p_0^\mu$ can be fully established in the case of the quadratic Hamiltonian in Eq.~\eqref{Hamiltonian} and extends to generic form of $g(q)$. 
A way to show this is goes as follows. Using the same techniques as outlined in the main text, it is very easy to show the saddle point equations for the fluctuating-$p_0$ model, can be written by  performing the formal substitution:
\beq
\int \de h P(q,h) \to \int \frac{\de \tilde p_0}{\sqrt{2\pi \sigma^2}} e^{-\tilde p_0^2/(2\s^2)}\int \de h P(q,h|\tilde p_0)
\eeq
and $P(q,h|\tilde p_0)$ satisfies exactly the same equations as in Eq.~\eqref{FP} but now with random initial condition
\beq
 P(q_m,h|\tilde p_0)=\frac{\exp\left[-(h+\tilde p_0)^2/(2g(q_m))\right]}{\sqrt{2\pi g(q_m)}}\:.
\eeq
This implies that $S(q)$ and its zero temperature limit $\hat S(q)$ are both proportional to $\tilde p_0$. Since all the saddle point equations are proportional to $S(q)^2$ and $\hat S(q)^2$ (see Eqs.~\eqref{SP1},\eqref{def_x} and \eqref{SP1_0}) averaging these quantities with the Gaussian distribution for $\tilde p_0$ gives a set of saddle point equations which coincide with the ones presented in the main text with $p_0^2$ replaced now by  $\sigma^2$. Note however that the shape index is not well defined for the fluctuating $p_0^\mu$ model since the average target shape is $\overline{p_0^\mu}=0$ and therefore the model with constant $p_0$ is the right one to compare with simulations of models of confluent tissues.}


\begin{thebibliography}{50}
\expandafter\ifx\csname natexlab\endcsname\relax\def\natexlab#1{#1}\fi
\expandafter\ifx\csname bibnamefont\endcsname\relax
  \def\bibnamefont#1{#1}\fi
\expandafter\ifx\csname bibfnamefont\endcsname\relax
  \def\bibfnamefont#1{#1}\fi
\expandafter\ifx\csname citenamefont\endcsname\relax
  \def\citenamefont#1{#1}\fi
\expandafter\ifx\csname url\endcsname\relax
  \def\url#1{\texttt{#1}}\fi
\expandafter\ifx\csname urlprefix\endcsname\relax\def\urlprefix{URL }\fi
\providecommand{\bibinfo}[2]{#2}
\providecommand{\eprint}[2][]{\url{#2}}

\bibitem[{\citenamefont{Sinha et~al.}(2020)\citenamefont{Sinha, Malmi-Kakkada,
  Li, Samanta, and Thirumalai}}]{sinha2020spatially}
\bibinfo{author}{\bibfnamefont{S.}~\bibnamefont{Sinha}},
  \bibinfo{author}{\bibfnamefont{A.~N.} \bibnamefont{Malmi-Kakkada}},
  \bibinfo{author}{\bibfnamefont{X.}~\bibnamefont{Li}},
  \bibinfo{author}{\bibfnamefont{H.~S.} \bibnamefont{Samanta}},
  \bibnamefont{and}
  \bibinfo{author}{\bibfnamefont{D.}~\bibnamefont{Thirumalai}},
  \bibinfo{journal}{Soft Matter} \textbf{\bibinfo{volume}{16}},
  \bibinfo{pages}{5294} (\bibinfo{year}{2020}).

\bibitem[{\citenamefont{Sinha and Thirumalai}(2020)}]{sinha2020self}
\bibinfo{author}{\bibfnamefont{S.}~\bibnamefont{Sinha}} \bibnamefont{and}
  \bibinfo{author}{\bibfnamefont{D.}~\bibnamefont{Thirumalai}},
  \bibinfo{journal}{The Journal of Chemical Physics}
  \textbf{\bibinfo{volume}{153}}, \bibinfo{pages}{201101}
  (\bibinfo{year}{2020}).

\bibitem[{\citenamefont{Blauth et~al.}(2021)\citenamefont{Blauth, Kubitschke,
  Gottheil, Grosser, and K{\"a}s}}]{blauth2021jamming}
\bibinfo{author}{\bibfnamefont{E.}~\bibnamefont{Blauth}},
  \bibinfo{author}{\bibfnamefont{H.}~\bibnamefont{Kubitschke}},
  \bibinfo{author}{\bibfnamefont{P.}~\bibnamefont{Gottheil}},
  \bibinfo{author}{\bibfnamefont{S.}~\bibnamefont{Grosser}}, \bibnamefont{and}
  \bibinfo{author}{\bibfnamefont{J.~A.} \bibnamefont{K{\"a}s}},
  \bibinfo{journal}{Frontiers in Physics} \textbf{\bibinfo{volume}{9}},
  \bibinfo{pages}{666709} (\bibinfo{year}{2021}).

\bibitem[{\citenamefont{Kandel et~al.}(2000)\citenamefont{Kandel, Schwartz,
  Jessell, Siegelbaum, Hudspeth, Mack et~al.}}]{kandel2000principles}
\bibinfo{author}{\bibfnamefont{E.~R.} \bibnamefont{Kandel}},
  \bibinfo{author}{\bibfnamefont{J.~H.} \bibnamefont{Schwartz}},
  \bibinfo{author}{\bibfnamefont{T.~M.} \bibnamefont{Jessell}},
  \bibinfo{author}{\bibfnamefont{S.}~\bibnamefont{Siegelbaum}},
  \bibinfo{author}{\bibfnamefont{A.~J.} \bibnamefont{Hudspeth}},
  \bibinfo{author}{\bibfnamefont{S.}~\bibnamefont{Mack}}, \bibnamefont{et~al.},
  \emph{\bibinfo{title}{Principles of neural science}},
  vol.~\bibinfo{volume}{4} (\bibinfo{publisher}{McGraw-hill New York},
  \bibinfo{year}{2000}).

\bibitem[{\citenamefont{Van~Essen}(2020)}]{van20202020}
\bibinfo{author}{\bibfnamefont{D.~C.} \bibnamefont{Van~Essen}},
  \bibinfo{journal}{Proceedings of the National Academy of Sciences}
  \textbf{\bibinfo{volume}{117}}, \bibinfo{pages}{32868}
  (\bibinfo{year}{2020}).

\bibitem[{\citenamefont{Tallinen et~al.}(2016)\citenamefont{Tallinen, Chung,
  Rousseau, Girard, Lef{\`e}vre, and Mahadevan}}]{tallinen2016growth}
\bibinfo{author}{\bibfnamefont{T.}~\bibnamefont{Tallinen}},
  \bibinfo{author}{\bibfnamefont{J.~Y.} \bibnamefont{Chung}},
  \bibinfo{author}{\bibfnamefont{F.}~\bibnamefont{Rousseau}},
  \bibinfo{author}{\bibfnamefont{N.}~\bibnamefont{Girard}},
  \bibinfo{author}{\bibfnamefont{J.}~\bibnamefont{Lef{\`e}vre}},
  \bibnamefont{and}
  \bibinfo{author}{\bibfnamefont{L.}~\bibnamefont{Mahadevan}},
  \bibinfo{journal}{Nature Physics} \textbf{\bibinfo{volume}{12}},
  \bibinfo{pages}{588} (\bibinfo{year}{2016}).

\bibitem[{\citenamefont{Bi et~al.}(2016)\citenamefont{Bi, Yang, Marchetti, and
  Manning}}]{bi2016motility}
\bibinfo{author}{\bibfnamefont{D.}~\bibnamefont{Bi}},
  \bibinfo{author}{\bibfnamefont{X.}~\bibnamefont{Yang}},
  \bibinfo{author}{\bibfnamefont{M.~C.} \bibnamefont{Marchetti}},
  \bibnamefont{and} \bibinfo{author}{\bibfnamefont{M.~L.}
  \bibnamefont{Manning}}, \bibinfo{journal}{Physical Review X}
  \textbf{\bibinfo{volume}{6}}, \bibinfo{pages}{021011} (\bibinfo{year}{2016}).

\bibitem[{\citenamefont{Bi et~al.}(2015)\citenamefont{Bi, Lopez, Schwarz, and
  Manning}}]{bi2015density}
\bibinfo{author}{\bibfnamefont{D.}~\bibnamefont{Bi}},
  \bibinfo{author}{\bibfnamefont{J.}~\bibnamefont{Lopez}},
  \bibinfo{author}{\bibfnamefont{J.~M.} \bibnamefont{Schwarz}},
  \bibnamefont{and} \bibinfo{author}{\bibfnamefont{M.~L.}
  \bibnamefont{Manning}}, \bibinfo{journal}{Nature Physics}
  \textbf{\bibinfo{volume}{11}}, \bibinfo{pages}{1074} (\bibinfo{year}{2015}).

\bibitem[{\citenamefont{Nagai and Honda}(2001)}]{nagai2001dynamic}
\bibinfo{author}{\bibfnamefont{T.}~\bibnamefont{Nagai}} \bibnamefont{and}
  \bibinfo{author}{\bibfnamefont{H.}~\bibnamefont{Honda}},
  \bibinfo{journal}{Philosophical Magazine B} \textbf{\bibinfo{volume}{81}},
  \bibinfo{pages}{699} (\bibinfo{year}{2001}).

\bibitem[{\citenamefont{Yan and Bi}(2019)}]{yan2019multicellular}
\bibinfo{author}{\bibfnamefont{L.}~\bibnamefont{Yan}} \bibnamefont{and}
  \bibinfo{author}{\bibfnamefont{D.}~\bibnamefont{Bi}},
  \bibinfo{journal}{Physical Review X} \textbf{\bibinfo{volume}{9}},
  \bibinfo{pages}{011029} (\bibinfo{year}{2019}).

\bibitem[{\citenamefont{Sussman and Merkel}(2018)}]{sussman2018no}
\bibinfo{author}{\bibfnamefont{D.~M.} \bibnamefont{Sussman}} \bibnamefont{and}
  \bibinfo{author}{\bibfnamefont{M.}~\bibnamefont{Merkel}},
  \bibinfo{journal}{Soft matter} \textbf{\bibinfo{volume}{14}},
  \bibinfo{pages}{3397} (\bibinfo{year}{2018}).

\bibitem[{\citenamefont{Merkel and Manning}(2018)}]{merkel2018geometrically}
\bibinfo{author}{\bibfnamefont{M.}~\bibnamefont{Merkel}} \bibnamefont{and}
  \bibinfo{author}{\bibfnamefont{M.~L.} \bibnamefont{Manning}},
  \bibinfo{journal}{New Journal of Physics} \textbf{\bibinfo{volume}{20}},
  \bibinfo{pages}{022002} (\bibinfo{year}{2018}).

\bibitem[{\citenamefont{Merkel et~al.}(2019)\citenamefont{Merkel, Baumgarten,
  Tighe, and Manning}}]{merkel2019minimal}
\bibinfo{author}{\bibfnamefont{M.}~\bibnamefont{Merkel}},
  \bibinfo{author}{\bibfnamefont{K.}~\bibnamefont{Baumgarten}},
  \bibinfo{author}{\bibfnamefont{B.~P.} \bibnamefont{Tighe}}, \bibnamefont{and}
  \bibinfo{author}{\bibfnamefont{M.~L.} \bibnamefont{Manning}},
  \bibinfo{journal}{Proceedings of the National Academy of Sciences}
  \textbf{\bibinfo{volume}{116}}, \bibinfo{pages}{6560} (\bibinfo{year}{2019}).

\bibitem[{\citenamefont{Parisi et~al.}(2020)\citenamefont{Parisi, Urbani, and
  Zamponi}}]{SimpleGlasses2020}
\bibinfo{author}{\bibfnamefont{G.}~\bibnamefont{Parisi}},
  \bibinfo{author}{\bibfnamefont{P.}~\bibnamefont{Urbani}}, \bibnamefont{and}
  \bibinfo{author}{\bibfnamefont{F.}~\bibnamefont{Zamponi}},
  \emph{\bibinfo{title}{Theory of simple glasses: exact solutions in infinite
  dimensions}} (\bibinfo{publisher}{Cambridge University Press},
  \bibinfo{year}{2020}).

\bibitem[{\citenamefont{Damavandi et~al.}(2022)\citenamefont{Damavandi, Hagh,
  Santangelo, and Manning}}]{damavandi2022energetic}
\bibinfo{author}{\bibfnamefont{O.~K.} \bibnamefont{Damavandi}},
  \bibinfo{author}{\bibfnamefont{V.~F.} \bibnamefont{Hagh}},
  \bibinfo{author}{\bibfnamefont{C.~D.} \bibnamefont{Santangelo}},
  \bibnamefont{and} \bibinfo{author}{\bibfnamefont{M.~L.}
  \bibnamefont{Manning}}, \bibinfo{journal}{Physical Review E}
  \textbf{\bibinfo{volume}{105}}, \bibinfo{pages}{025004}
  (\bibinfo{year}{2022}).

\bibitem[{\citenamefont{Sussman et~al.}(2018)\citenamefont{Sussman, Paoluzzi,
  Marchetti, and Manning}}]{sussman2018anomalous}
\bibinfo{author}{\bibfnamefont{D.~M.} \bibnamefont{Sussman}},
  \bibinfo{author}{\bibfnamefont{M.}~\bibnamefont{Paoluzzi}},
  \bibinfo{author}{\bibfnamefont{M.~C.} \bibnamefont{Marchetti}},
  \bibnamefont{and} \bibinfo{author}{\bibfnamefont{M.~L.}
  \bibnamefont{Manning}}, \bibinfo{journal}{EPL (Europhysics Letters)}
  \textbf{\bibinfo{volume}{121}}, \bibinfo{pages}{36001}
  (\bibinfo{year}{2018}).

\bibitem[{\citenamefont{Franz and Parisi}(2016)}]{franz2016simplest}
\bibinfo{author}{\bibfnamefont{S.}~\bibnamefont{Franz}} \bibnamefont{and}
  \bibinfo{author}{\bibfnamefont{G.}~\bibnamefont{Parisi}},
  \bibinfo{journal}{Journal of Physics A: Mathematical and Theoretical}
  \textbf{\bibinfo{volume}{49}}, \bibinfo{pages}{145001}
  (\bibinfo{year}{2016}).

\bibitem[{\citenamefont{Franz et~al.}(2017)\citenamefont{Franz, Parisi,
  Sevelev, Urbani, and Zamponi}}]{FPSUZ17}
\bibinfo{author}{\bibfnamefont{S.}~\bibnamefont{Franz}},
  \bibinfo{author}{\bibfnamefont{G.}~\bibnamefont{Parisi}},
  \bibinfo{author}{\bibfnamefont{M.}~\bibnamefont{Sevelev}},
  \bibinfo{author}{\bibfnamefont{P.}~\bibnamefont{Urbani}}, \bibnamefont{and}
  \bibinfo{author}{\bibfnamefont{F.}~\bibnamefont{Zamponi}},
  \bibinfo{journal}{SciPost Physics} \textbf{\bibinfo{volume}{2}},
  \bibinfo{pages}{019} (\bibinfo{year}{2017}).

\bibitem[{\citenamefont{Brito et~al.}(2018)\citenamefont{Brito, Ikeda, Urbani,
  Wyart, and Zamponi}}]{brito2018universality}
\bibinfo{author}{\bibfnamefont{C.}~\bibnamefont{Brito}},
  \bibinfo{author}{\bibfnamefont{H.}~\bibnamefont{Ikeda}},
  \bibinfo{author}{\bibfnamefont{P.}~\bibnamefont{Urbani}},
  \bibinfo{author}{\bibfnamefont{M.}~\bibnamefont{Wyart}}, \bibnamefont{and}
  \bibinfo{author}{\bibfnamefont{F.}~\bibnamefont{Zamponi}},
  \bibinfo{journal}{Proceedings of the National Academy of Sciences}
  \textbf{\bibinfo{volume}{115}}, \bibinfo{pages}{11736}
  (\bibinfo{year}{2018}).

\bibitem[{\citenamefont{Ikeda et~al.}(2019)\citenamefont{Ikeda, Urbani, and
  Zamponi}}]{ikeda2019mean}
\bibinfo{author}{\bibfnamefont{H.}~\bibnamefont{Ikeda}},
  \bibinfo{author}{\bibfnamefont{P.}~\bibnamefont{Urbani}}, \bibnamefont{and}
  \bibinfo{author}{\bibfnamefont{F.}~\bibnamefont{Zamponi}},
  \bibinfo{journal}{Journal of Physics A: Mathematical and Theoretical}
  \textbf{\bibinfo{volume}{52}}, \bibinfo{pages}{344001}
  (\bibinfo{year}{2019}).

\bibitem[{\citenamefont{Franz et~al.}(2019)\citenamefont{Franz, Sclocchi, and
  Urbani}}]{FSU19}
\bibinfo{author}{\bibfnamefont{S.}~\bibnamefont{Franz}},
  \bibinfo{author}{\bibfnamefont{A.}~\bibnamefont{Sclocchi}}, \bibnamefont{and}
  \bibinfo{author}{\bibfnamefont{P.}~\bibnamefont{Urbani}},
  \bibinfo{journal}{Physical review letters} \textbf{\bibinfo{volume}{123}},
  \bibinfo{pages}{115702} (\bibinfo{year}{2019}).

\bibitem[{\citenamefont{Sclocchi and Urbani}(2021)}]{SU21}
\bibinfo{author}{\bibfnamefont{A.}~\bibnamefont{Sclocchi}} \bibnamefont{and}
  \bibinfo{author}{\bibfnamefont{P.}~\bibnamefont{Urbani}},
  \bibinfo{journal}{SciPost Phys.} \textbf{\bibinfo{volume}{10}},
  \bibinfo{pages}{13} (\bibinfo{year}{2021}),
  \urlprefix\url{https://scipost.org/10.21468/SciPostPhys.10.1.013}.

\bibitem[{\citenamefont{Sclocchi and Urbani}(2022)}]{sclocchi2022high}
\bibinfo{author}{\bibfnamefont{A.}~\bibnamefont{Sclocchi}} \bibnamefont{and}
  \bibinfo{author}{\bibfnamefont{P.}~\bibnamefont{Urbani}},
  \bibinfo{journal}{Physical Review E} \textbf{\bibinfo{volume}{105}},
  \bibinfo{pages}{024134} (\bibinfo{year}{2022}).

\bibitem[{\citenamefont{Charbonneau et~al.}(2014)\citenamefont{Charbonneau,
  Kurchan, Parisi, Urbani, and Zamponi}}]{CKPUZ14}
\bibinfo{author}{\bibfnamefont{P.}~\bibnamefont{Charbonneau}},
  \bibinfo{author}{\bibfnamefont{J.}~\bibnamefont{Kurchan}},
  \bibinfo{author}{\bibfnamefont{G.}~\bibnamefont{Parisi}},
  \bibinfo{author}{\bibfnamefont{P.}~\bibnamefont{Urbani}}, \bibnamefont{and}
  \bibinfo{author}{\bibfnamefont{F.}~\bibnamefont{Zamponi}},
  \bibinfo{journal}{Nature communications} \textbf{\bibinfo{volume}{5}},
  \bibinfo{pages}{1} (\bibinfo{year}{2014}).

\bibitem[{\citenamefont{Charbonneau et~al.}(2021)\citenamefont{Charbonneau,
  Corwin, Dennis, Rojas, Ikeda, Parisi, and
  Ricci-Tersenghi}}]{charbonneau2021finite}
\bibinfo{author}{\bibfnamefont{P.}~\bibnamefont{Charbonneau}},
  \bibinfo{author}{\bibfnamefont{E.~I.} \bibnamefont{Corwin}},
  \bibinfo{author}{\bibfnamefont{R.~C.} \bibnamefont{Dennis}},
  \bibinfo{author}{\bibfnamefont{R.~D.~H.} \bibnamefont{Rojas}},
  \bibinfo{author}{\bibfnamefont{H.}~\bibnamefont{Ikeda}},
  \bibinfo{author}{\bibfnamefont{G.}~\bibnamefont{Parisi}}, \bibnamefont{and}
  \bibinfo{author}{\bibfnamefont{F.}~\bibnamefont{Ricci-Tersenghi}},
  \bibinfo{journal}{Physical Review E} \textbf{\bibinfo{volume}{104}},
  \bibinfo{pages}{014102} (\bibinfo{year}{2021}).

\bibitem[{\citenamefont{Franz et~al.}(2020)\citenamefont{Franz, Sclocchi, and
  Urbani}}]{franz2020critical}
\bibinfo{author}{\bibfnamefont{S.}~\bibnamefont{Franz}},
  \bibinfo{author}{\bibfnamefont{A.}~\bibnamefont{Sclocchi}}, \bibnamefont{and}
  \bibinfo{author}{\bibfnamefont{P.}~\bibnamefont{Urbani}},
  \bibinfo{journal}{SciPost Physics} \textbf{\bibinfo{volume}{9}},
  \bibinfo{pages}{012} (\bibinfo{year}{2020}).

\bibitem[{\citenamefont{Agoritsas et~al.}(2018)\citenamefont{Agoritsas, Biroli,
  Urbani, and Zamponi}}]{agoritsas2018out}
\bibinfo{author}{\bibfnamefont{E.}~\bibnamefont{Agoritsas}},
  \bibinfo{author}{\bibfnamefont{G.}~\bibnamefont{Biroli}},
  \bibinfo{author}{\bibfnamefont{P.}~\bibnamefont{Urbani}}, \bibnamefont{and}
  \bibinfo{author}{\bibfnamefont{F.}~\bibnamefont{Zamponi}},
  \bibinfo{journal}{Journal of Physics A: Mathematical and Theoretical}
  \textbf{\bibinfo{volume}{51}}, \bibinfo{pages}{085002}
  (\bibinfo{year}{2018}).

\bibitem[{\citenamefont{Manacorda and Zamponi}(2022)}]{Manacorda_2022}
\bibinfo{author}{\bibfnamefont{A.}~\bibnamefont{Manacorda}} \bibnamefont{and}
  \bibinfo{author}{\bibfnamefont{F.}~\bibnamefont{Zamponi}},
  \bibinfo{journal}{Journal of Physics A: Mathematical and Theoretical}
  \textbf{\bibinfo{volume}{55}}, \bibinfo{pages}{334001}
  (\bibinfo{year}{2022}),
  \urlprefix\url{https://dx.doi.org/10.1088/1751-8121/ac7f06}.

\bibitem[{\citenamefont{Mignacco et~al.}(2020)\citenamefont{Mignacco, Krzakala,
  Urbani, and Zdeborov{\'a}}}]{mignacco2020dynamical}
\bibinfo{author}{\bibfnamefont{F.}~\bibnamefont{Mignacco}},
  \bibinfo{author}{\bibfnamefont{F.}~\bibnamefont{Krzakala}},
  \bibinfo{author}{\bibfnamefont{P.}~\bibnamefont{Urbani}}, \bibnamefont{and}
  \bibinfo{author}{\bibfnamefont{L.}~\bibnamefont{Zdeborov{\'a}}},
  \bibinfo{journal}{Advances in Neural Information Processing Systems}
  \textbf{\bibinfo{volume}{33}}, \bibinfo{pages}{9540} (\bibinfo{year}{2020}).

\bibitem[{\citenamefont{Fyodorov}(2019)}]{fyodorov2019spin}
\bibinfo{author}{\bibfnamefont{Y.~V.} \bibnamefont{Fyodorov}},
  \bibinfo{journal}{Journal of Statistical Physics}
  \textbf{\bibinfo{volume}{175}}, \bibinfo{pages}{789} (\bibinfo{year}{2019}).

\bibitem[{\citenamefont{Fyodorov and Tublin}(2022)}]{fyodorov2022optimization}
\bibinfo{author}{\bibfnamefont{Y.~V.} \bibnamefont{Fyodorov}} \bibnamefont{and}
  \bibinfo{author}{\bibfnamefont{R.}~\bibnamefont{Tublin}},
  \bibinfo{journal}{Journal of Physics A: Mathematical and Theoretical}
  \textbf{\bibinfo{volume}{55}}, \bibinfo{pages}{244008}
  (\bibinfo{year}{2022}).

\bibitem[{\citenamefont{Ikeda}(2022)}]{ikeda2022sat}
\bibinfo{author}{\bibfnamefont{H.}~\bibnamefont{Ikeda}},
  \bibinfo{journal}{arXiv preprint arXiv:2208.08162}  (\bibinfo{year}{2022}).

\bibitem[{\citenamefont{Huang et~al.}(2020)\citenamefont{Huang, Gupta, and
  Dokmani{\'c}}}]{huang2020solving}
\bibinfo{author}{\bibfnamefont{S.}~\bibnamefont{Huang}},
  \bibinfo{author}{\bibfnamefont{S.}~\bibnamefont{Gupta}}, \bibnamefont{and}
  \bibinfo{author}{\bibfnamefont{I.}~\bibnamefont{Dokmani{\'c}}},
  \bibinfo{journal}{IEEE Transactions on Signal Processing}
  \textbf{\bibinfo{volume}{68}}, \bibinfo{pages}{4782} (\bibinfo{year}{2020}).

\bibitem[{\citenamefont{Tublin}(2022)}]{tublin2022few}
\bibinfo{author}{\bibfnamefont{R.}~\bibnamefont{Tublin}}, Ph.D. thesis
  (\bibinfo{year}{2022}).

\bibitem[{\citenamefont{M{\'e}zard and Parisi}(1991)}]{mezard1991replica}
\bibinfo{author}{\bibfnamefont{M.}~\bibnamefont{M{\'e}zard}} \bibnamefont{and}
  \bibinfo{author}{\bibfnamefont{G.}~\bibnamefont{Parisi}},
  \bibinfo{journal}{Journal de Physique I} \textbf{\bibinfo{volume}{1}},
  \bibinfo{pages}{809} (\bibinfo{year}{1991}).

\bibitem[{\citenamefont{Mezard et~al.}(1987)\citenamefont{Mezard, Parisi, and
  Virasoro}}]{MPV87}
\bibinfo{author}{\bibfnamefont{M.}~\bibnamefont{Mezard}},
  \bibinfo{author}{\bibfnamefont{G.}~\bibnamefont{Parisi}}, \bibnamefont{and}
  \bibinfo{author}{\bibfnamefont{M.~A.} \bibnamefont{Virasoro}},
  \emph{\bibinfo{title}{Spin glass theory and beyond}}
  (\bibinfo{publisher}{World Scientific}, \bibinfo{address}{Singapore},
  \bibinfo{year}{1987}).

\bibitem[{\citenamefont{Parisi}(1980)}]{Pa80}
\bibinfo{author}{\bibfnamefont{G.}~\bibnamefont{Parisi}},
  \bibinfo{journal}{Journal of Physics A: Mathematical and General}
  \textbf{\bibinfo{volume}{13}}, \bibinfo{pages}{L115} (\bibinfo{year}{1980}).

\bibitem[{\citenamefont{Sommers and Dupont}(1984)}]{SD84}
\bibinfo{author}{\bibfnamefont{H.-J.} \bibnamefont{Sommers}} \bibnamefont{and}
  \bibinfo{author}{\bibfnamefont{W.}~\bibnamefont{Dupont}},
  \bibinfo{journal}{Journal of Physics C: Solid State Physics}
  \textbf{\bibinfo{volume}{17}}, \bibinfo{pages}{5785} (\bibinfo{year}{1984}).

\bibitem[{\citenamefont{de~Almeida and Thouless}(1978)}]{de1978stability}
\bibinfo{author}{\bibfnamefont{J.~R.} \bibnamefont{de~Almeida}}
  \bibnamefont{and} \bibinfo{author}{\bibfnamefont{D.~J.}
  \bibnamefont{Thouless}}, \bibinfo{journal}{Journal of Physics A: Mathematical
  and General} \textbf{\bibinfo{volume}{11}}, \bibinfo{pages}{983}
  (\bibinfo{year}{1978}).

\bibitem[{\citenamefont{Sommers}(1985)}]{sommers1985parisi}
\bibinfo{author}{\bibfnamefont{H.-J.} \bibnamefont{Sommers}},
  \bibinfo{journal}{Journal de Physique Lettres} \textbf{\bibinfo{volume}{46}},
  \bibinfo{pages}{779} (\bibinfo{year}{1985}).

\bibitem[{\citenamefont{Kurchan et~al.}(2013)\citenamefont{Kurchan, Parisi,
  Urbani, and Zamponi}}]{KPUZ13}
\bibinfo{author}{\bibfnamefont{J.}~\bibnamefont{Kurchan}},
  \bibinfo{author}{\bibfnamefont{G.}~\bibnamefont{Parisi}},
  \bibinfo{author}{\bibfnamefont{P.}~\bibnamefont{Urbani}}, \bibnamefont{and}
  \bibinfo{author}{\bibfnamefont{F.}~\bibnamefont{Zamponi}},
  \bibinfo{journal}{J. Phys. Chem. B} \textbf{\bibinfo{volume}{117}},
  \bibinfo{pages}{12979} (\bibinfo{year}{2013}),
  \urlprefix\url{https://pubs.acs.org/doi/10.1021/jp402235d}.

\bibitem[{\citenamefont{Charbonneau et~al.}(2017)\citenamefont{Charbonneau,
  Kurchan, Parisi, Urbani, and Zamponi}}]{CKPUZ17}
\bibinfo{author}{\bibfnamefont{P.}~\bibnamefont{Charbonneau}},
  \bibinfo{author}{\bibfnamefont{J.}~\bibnamefont{Kurchan}},
  \bibinfo{author}{\bibfnamefont{G.}~\bibnamefont{Parisi}},
  \bibinfo{author}{\bibfnamefont{P.}~\bibnamefont{Urbani}}, \bibnamefont{and}
  \bibinfo{author}{\bibfnamefont{F.}~\bibnamefont{Zamponi}},
  \bibinfo{journal}{Annual Review of Condensed Matter Physics}
  \textbf{\bibinfo{volume}{8}}, \bibinfo{pages}{265} (\bibinfo{year}{2017}).

\bibitem[{\citenamefont{Urbani and Biroli}(2015)}]{BU15}
\bibinfo{author}{\bibfnamefont{P.}~\bibnamefont{Urbani}} \bibnamefont{and}
  \bibinfo{author}{\bibfnamefont{G.}~\bibnamefont{Biroli}},
  \bibinfo{journal}{Phys. Rev. B} \textbf{\bibinfo{volume}{91}},
  \bibinfo{pages}{100202} (\bibinfo{year}{2015}).

\bibitem[{\citenamefont{Pinto et~al.}(2022)\citenamefont{Pinto, Sussman,
  da~Gama, and Ara{\'u}jo}}]{pinto2022hierarchical}
\bibinfo{author}{\bibfnamefont{D.~E.} \bibnamefont{Pinto}},
  \bibinfo{author}{\bibfnamefont{D.~M.} \bibnamefont{Sussman}},
  \bibinfo{author}{\bibfnamefont{M.~M.~T.} \bibnamefont{da~Gama}},
  \bibnamefont{and} \bibinfo{author}{\bibfnamefont{N.~A.}
  \bibnamefont{Ara{\'u}jo}}, \bibinfo{journal}{Physical Review Research}
  \textbf{\bibinfo{volume}{4}}, \bibinfo{pages}{023187} (\bibinfo{year}{2022}).

\bibitem[{\citenamefont{Li et~al.}(2021)\citenamefont{Li, Wei, Paoluzzi, and
  Ciamarra}}]{li2021softness}
\bibinfo{author}{\bibfnamefont{Y.-W.} \bibnamefont{Li}},
  \bibinfo{author}{\bibfnamefont{L.~L.~Y.} \bibnamefont{Wei}},
  \bibinfo{author}{\bibfnamefont{M.}~\bibnamefont{Paoluzzi}}, \bibnamefont{and}
  \bibinfo{author}{\bibfnamefont{M.~P.} \bibnamefont{Ciamarra}},
  \bibinfo{journal}{Physical Review E} \textbf{\bibinfo{volume}{103}},
  \bibinfo{pages}{022607} (\bibinfo{year}{2021}).

\bibitem[{\citenamefont{Franz et~al.}(2015)\citenamefont{Franz, Parisi, Urbani,
  and Zamponi}}]{FPUZ15}
\bibinfo{author}{\bibfnamefont{S.}~\bibnamefont{Franz}},
  \bibinfo{author}{\bibfnamefont{G.}~\bibnamefont{Parisi}},
  \bibinfo{author}{\bibfnamefont{P.}~\bibnamefont{Urbani}}, \bibnamefont{and}
  \bibinfo{author}{\bibfnamefont{F.}~\bibnamefont{Zamponi}},
  \bibinfo{journal}{Proc. Natl. Acad. Sci. U.S.A.}
  \textbf{\bibinfo{volume}{112}}, \bibinfo{pages}{14539}
  (\bibinfo{year}{2015}).

\bibitem[{\citenamefont{Urbani}(2022)}]{urbani2022field}
\bibinfo{author}{\bibfnamefont{P.}~\bibnamefont{Urbani}},
  \bibinfo{journal}{Journal of Physics A: Mathematical and Theoretical}
  \textbf{\bibinfo{volume}{55}}, \bibinfo{pages}{335002}
  (\bibinfo{year}{2022}).

\bibitem[{\citenamefont{Lerner and Bouchbinder}(2021)}]{lerner2021low}
\bibinfo{author}{\bibfnamefont{E.}~\bibnamefont{Lerner}} \bibnamefont{and}
  \bibinfo{author}{\bibfnamefont{E.}~\bibnamefont{Bouchbinder}},
  \bibinfo{journal}{The Journal of chemical physics}
  \textbf{\bibinfo{volume}{155}}, \bibinfo{pages}{200901}
  (\bibinfo{year}{2021}).

\bibitem[{\citenamefont{Bouchbinder et~al.}(2021)\citenamefont{Bouchbinder,
  Lerner, Rainone, Urbani, and Zamponi}}]{bouchbinder2021low}
\bibinfo{author}{\bibfnamefont{E.}~\bibnamefont{Bouchbinder}},
  \bibinfo{author}{\bibfnamefont{E.}~\bibnamefont{Lerner}},
  \bibinfo{author}{\bibfnamefont{C.}~\bibnamefont{Rainone}},
  \bibinfo{author}{\bibfnamefont{P.}~\bibnamefont{Urbani}}, \bibnamefont{and}
  \bibinfo{author}{\bibfnamefont{F.}~\bibnamefont{Zamponi}},
  \bibinfo{journal}{Physical Review B} \textbf{\bibinfo{volume}{103}},
  \bibinfo{pages}{174202} (\bibinfo{year}{2021}).

\bibitem[{\citenamefont{Galassi}(3rd Ed.)}]{galassi2002gnu}
\bibinfo{author}{\bibfnamefont{M.~e.~a.} \bibnamefont{Galassi}},
  \emph{\bibinfo{title}{GNU Scientific Library Reference Manual}}
  (\bibinfo{year}{3rd Ed.}).

\end{thebibliography}

\end{document}